\documentstyle[twocolumn,aps]{revtex}
\begin{document}

\draft
\title{Quasiparticles in Finite-Temperature Field Theory}
\author{H. Arthur Weldon}
\address{Department of Physics, West Virginia University, Morgantown West
Virginia 26506-6315}
\date{September 10, 1998}
\maketitle

\begin{abstract}
Conventional finite-temperature  perturbation theory in which  propagators have poles at $k^{2}=m^{2}$ is shown to
break down  at the two-loop level for self-interacting scalar fields.
The breakdown is avoided by using free thermal propagators that have poles at
the same energy as the exact thermal propagator. 
This quasiparticle energy ${\cal E}(\vec{k})$  is 
temperature-dependent, complex, and gauge invariant.
An operator theory containing two self-adjoint scalar fields
is presented in which  all temperature
dependence is incorporated into the Hamiltonian. 
No thermal traces are required to compute thermal Green functions. Choosing
the spectrum of the unperturbed part of the Hamiltonian to contain the exact
quasiparticle energy ${\cal E}(\vec{k})$ produces a resummed perturbation theory
that has the correct poles and branch cuts. The location of the poles and cuts
is explained directly in terms of the spectrum of the Hamiltonian.
\end{abstract}


\section{Introduction}

In high temperature QCD the self-energies of both  quarks and gluons
change the particle dispersion relations from their
zero-temperature form. The modified dispersion relations indicate
that the naturally propagating modes are collective
excitations, i.e. quasiparticles.
The locations of the poles in the propagators are gauge-fixing
invariant \cite{KKR}.   Braaten and Pisarski showed that for  a consistent
perturbation theory the propagators should not be free but should include
part of the self-energies \cite{BP}. 
 With this hard-thermal-loop approximation to the quasiparticle effects it is 
possible to compute gauge invariant rates in which the corrections are
controllably small, at least in certain kinematic regimes. A completely
different test of the HTL quasiparticle dispersion relations comes from the
excellent agreement with lattice data that is obtained by using an ideal gas of
quark and gluon quasiparticles. When running coupling constants are
incorporated, the agreement holds even very close to the critical temperature of
QCD \cite{lat}.

Quasiparticles are ubiquitous in finite-temperature calculations, and yet 
they normally play no role in the operator
structure of the  theory. This situation contrasts markedly with 
that of  conventional particles at T=0. In particular, zero-temperature
field theory has the following simple properties:
\begin{itemize}
\item  The Hamiltonian operator has eigenstates $|\Psi\rangle$ in a
linear vector space.
\item Poles in the propagator occur at the eigenvalues of the 
single particle state vectors.
\item Branch cuts in the propagator occur at the eigenvalues of
the multiparticle state vectors.
\end{itemize}
\noindent This paper will describe a formalism in which these features
also hold for finite-temperature quasiparticles.
The present discussion will only treat scalar field theory. 

Sections II and III summarize the results of a recent paper on branch cuts 
at finite temperature \cite{AW1}, which shows that the location of the cuts is
determined by the quasiparticle energies. Section IV is based on a
new formulation of finite-temperature field theory \cite{AW2} in which
all the Bose-Einstein temperature dependence is contained in the Lagrangian.
Green functions are computed as vacuum matrix elements rather than thermal
traces. Sections V and VI show how to split the Lagrangian into an unperturbed
part that  describes free quasiparticles and a remainder that determines the
interactions. Several illustrations are worked out. The developments in Sec. IV
and subsequently may be read independently of the earlier sections.

\section{Breakdown of perturbation theory in massive $\phi^{3}$ field theory}

Braaten and Pisarski showed that finite-temperature  perturbation theory
breaks down in 
field theories with massless particles or those in which the mass is much
smaller than the temperature. Resummation of  hard thermal loops is then
necessary for Green functions evaluated at low energy and momentum.

This section will discuss a more primitive breakdown of perturbation
theory that occurs even without soft external momenta and without high
temperatures. The discussion will treat  scalar field theory having cubic
self-interactions,
\begin{equation}
{\cal L}={1\over 2}\partial_{\mu}\phi\partial^{\mu}\phi
-{1\over 2} m^{2}\phi^{2}- {g\over 3!}\phi^{3},\end{equation}
with counterterms omitted.
Perturbation theory is conventionally defined by 
choosing free thermal propagators that have poles at  the zero-temperature
 mass $m$. This produces two-particle branch points in the self-energy at
$k^{2}=4m^{2}$ and three-particle cuts at $k^{2}=9m^{2}$.
Simple calculations show that at finite temperature perturbation theory fails in
the vicinity of these points. The reason is that they are not the true branch
points of the full theory.

\subsection{Zero-temperature example}

A simple zero-temperature example will illustrate how higher order corrections
can shift the location of branch cuts.
Even though $m$ is the physical mass for the theory, suppose  that one performs
perturbative calculations using a free propagator
$\Delta(k)=1/[k^{2}-m_{0}^{2}]$, where
$m_{0}$ is some different mass.  It is important that $m_{0}$ is  finite and
not the bare mass. For definiteness choose $m_{0}<m$. The 
one-loop self-energy shown in Fig. 1 has 
a branch cut for $k^{2}\ge 4m_{0}^{2}$. The discontinuity across the branch cut
is
\begin{equation}
{\rm Disc}\,\Pi^{(1)}(k)=
{-ig^{2}\over 16\pi}\big(1-{4m_{0}^{2}\over k^{2}}\big)^{1/2}
\;\theta(k^{2}-4m_{0}^{2}).
\end{equation}
Since $m_{0}$ has no physical significance, this is clearly not a branch cut of
the full theory.
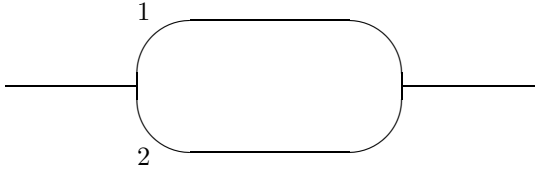
\begin{figure}
\begin{picture}(200,100)(-10,0)
\put(100,50){\oval(100,50)}
\put(0,50){\line(1,0){50}}
\put(150,50){\line(1,0){50}}
\put(50,75){1}
\put(50,20){2}
\end{picture}
\caption{ One-loop self-energy.}\label{fig1}
\end{figure}
The indication that $k^{2}=4m_{0}^{2}$ is not a branch point of the full theory
comes from the two-loop contribution. The full propagator is
$D'(k)=1/[k^{2}-m^{2}-\Pi(k)]$ and by  definition $\Pi$ contains the necessary
counterterm to vanish at the true mass $k^{2}=m^{2}$. To do perturbation theory
with mass $m_{0}$ the full propagator is written
$D'(k)=1/[k^{2}-m_{0}^{2}-\tilde{\Pi}(k)]$ where 
$\tilde{\Pi}(k)=m^{2}-m_{0}^{2}+\Pi(k)$. Of course $\tilde{\Pi}$ does not
vanish at
$k^{2}=m^{2}$ or at $k^{2}=m_{0}^{2}$ and this is the source of the problem.  
A self-energy insertion on the internal lines of Fig. 1 gives the 
two-loop contribution shown in Fig. 2.
\begin{figure}
\begin{picture}(200,110)(-10,0)
\put(100,50){\oval(100,50)}
\put(100,75){\oval(60,40)[t]}
\put(0,50){\line(1,0){50}}
\put(150,50){\line(1,0){50}}
\put(45,70){1}
\put(150,70){1}
\put(100,14){2}
\put(100,65){3}
\put(100,85){4}
\end{picture}
\caption{Two-loop self-energy due to one  insertion.}
\label{fig2}
\end{figure}
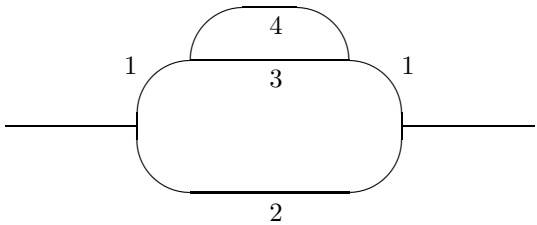
The corresponding  self-energy is
\begin{equation}
\Pi^{(2)}(k)=ig^{2}\int{d^{4}p\over (2\pi)^{4}}\big[\Delta(p)\tilde{\Pi}(p)\Delta(p)\big]
\Delta(p-k).\end{equation}
This has a two-particle and a three-particle discontinuity. The quantity in
square brackets has a double pole at $p^{2}=m_{0}^{2}$ since $\tilde{\Pi}(p)$
does not vanish at $p^{2}=m_{0}^{2}$.   This produces a  two-particle
discontinuity of the form 
\begin{displaymath}
{\rm Disc}\,\Pi^{(2)}(k)\!=\!{ig^{2}\over 8\pi}
\Big[{\delta m^{2}/k^{2}\over
\big(1-4m_{0}^{2}/ k^{2}\big)^{1/2}}
+\dots\Big]\theta(k^{2}-4m_{0}^{2}).
\end{displaymath}
The correction is
is infinite at  $k^{2}=4m_{0}^{2}$ and large near there. This
 is a signal that the correct branch point is not at $k^{2}=4m_{0}^{2}$.
 Multiple  self-energy insertions   produce successively higher powers of the
inverse square root:
\begin{eqnarray}
{\rm Disc}\,\Pi(k)=&&{-ig^{2}\over 16\pi}
\Big[\big(1-{4m_{0}^{2}\over k^{2}}\big)^{1/2}
-{2\delta m^{2}\over k^{2}}\big(1-{4m_{0}^{2}\over k^{2}}\big)^{-1/2}
\nonumber\\
&&\hskip0.5cm -{1\over 2}\big({2\delta m^{2}\over k^{2}}\big)^{2}
\big(1-{4m_{0}^{2}\over k^{2}}\big)^{-3/2}+\dots\Big].\nonumber\end{eqnarray}
This is the beginning of a Taylor series. All the corrections diverge at
the false threshold $k^{2}=4m_{0}^{2}$. In the range
$4m_{0}^{2}<k^{2}<4m^{2}$ each correction is finite but the
Taylor series diverges. Thus perturbation theory fails throughout this region of
$k^{2}$. To obtain a convergent series it is necessary to work in the range
$k^{2}>4m^{2}$.
In this region the  Taylor series converges
 and the sum is
the full two-particle discontinuity with branch point shifted to the physical mass
$m^{2}=m_{0}^{2}+\delta m^{2}$:
\begin{equation}
 {\rm Disc}\,\Pi(k)
={-ig^{2}\over 16\pi}\big(1-{4m^{2}\over k^{2}})^{1/2}\theta(k^{2}-4m^{2}).
\end{equation}
The true two-particle threshold is still a square root branch point at
$k^{2}=4m^{2}$. 
The breakdown of perturbation theory is entirely due to a propagator
$\Delta(k)=1/[k^{2}-m_{0}^{2}]$ with the the wrong mass $m_{0}$. The breakdown is easily
avoided by using $1/[k^{2}-m^{2}]$ for the free particle propagator.

\subsection{Breakdown of thermal perturbation theory}

In the
preceding example $m_{0}$ was the wrong mass. Finite temperature calculations
are normally done using  free propagators  with the wrong mass, i.e. the 
zero-temperature mass $m$. A typical component of the conventional
finite-temperature matrix propagator 
$\Delta_{ab}(k)$ is
\begin{displaymath}
\Delta_{++}(k)={1\over k^{2}-m^{2}+i\epsilon}-{2\pi i\delta(k^{2}-m^{2})
\over \exp(\beta|k_{0}|)-1}.\end{displaymath}
The two-loop topology shown in Fig. 2 produces a matrix
self-energy
\begin{displaymath}
\Pi_{ad}^{(2)}(k)=ig^{2}\int {d^{4}p\over (2\pi)^{4}}
\big[\Delta_{ab}(p)\Pi^{(1)}_{bc}(p)\Delta_{cd}(p)\big]\Delta_{ad}(k-p).
\end{displaymath}
As in Eq. (3) the quantity in square brackets has a double pole at
$p^{2}=m^{2}$ and  is the source of the problem. The self-energy has branch
cuts at two-particle and three-particle thresholds as well as cuts due to
absorption of one or more particles from the heat bath. The discontinuities
across any of these is given by the cutting rules of Kobes and Semenoff
\cite{KS}. A particular linear combination of the $\Pi_{ad}$ gives the retarded
self-energy. The formula for the discontinuity of the two-loop retarded
self-energy is given explicitly in
\cite{LB,Gel} in terms of the one-loop
$\Pi^{(1)}_{R}$. It is easily evaluated in the limit $k^{2}\to 4m^{2}$
with the result  
\begin{displaymath}
{\rm Disc}\,\Pi_{R}^{(2)}(k)\!\to {-ig^{2}\over 16\pi}{ [1+2n(k_{0}/2)]
{\rm Re}\,\Pi_{R}^{(1)}(k_{0}/2)\over
k^{2}(1-4m^{2}/k^{2})^{1/2}}.\end{displaymath}
The correction is infinite at $k^{2}=4m^{2}$ indicating the breakdown of
 finite-temperature perturbation theory in this region.
This breakdown is an artifact of using propagators with mass $m$. 

\section{Quasiparticle Resummation for $\phi^{3}$ field theory}

To cure the breakdown of thermal perturbation theory described above it is
necessary to use a free thermal propagator that has the same poles
as the exact thermal propagator. 
The exact thermal propagator matrix may be expressed as a linear
combination of the exact retarded and advanced thermal propagators.
The exact retarded thermal propagator $D_{R}'(k)$ has the following properties:
(a) It is analytic for Im $k_{0}>0$;
(b) It has a pole in the fourth quadrant at $k_{0}={\cal E}(\vec{k})$; 
(c) It has
a pole in the third quadrant at $k_{0}=-{\cal E}^{*}(\vec{k})$
because of the relation
$D_{R}'(k_{0},\vec{k})=\big[D_{R}'(-k_{0}^{*},\vec{k})\big]^{*}$.  It will be
useful to have an explicit notation for the real and imaginary parts of
${\cal E}$:
\begin{equation}
{\cal E}(\vec{k})=E(\vec{k})-i\Gamma(\vec{k})/2\hskip1cm (E>0;
\Gamma>0).\end{equation}
The fourth-quadrant pole  is thus at $-{\cal
E}^{*}=-E-i\Gamma/2$. The functions $E$ and $\Gamma$
also depend on $T, g, m$. 

It is  somewhat misleading to write the retarded propagator as
$D_{R}'(k)=1/[k^{2}-m^{2}-\Pi_{R}(k)]$
 since there is no pole at $k^{2}=m^{2}$. It is better to express it
as
\begin{displaymath}
D_{R}'(k)={1\over(k_{0}-{\cal E})(k_{0}+{\cal
E}^{*})-\Pi_{Rqp}(k)},\end{displaymath}
where $\Pi_{Rqp}(k)=\Pi_{R}(k)+\vec{k}^{2}+m^{2}-|{\cal E}|^{2}+i\Gamma k_{0}$.
It is natural to treat $\Pi_{Rqp}$ perturbatively since in vanishes at
$k_{0}={\cal E}$ and at $k_{0}=-{\cal E}^{*}$. Consequently the free retarded
quasiparticle propagator is 
\begin{equation}
D_{R}(k)={1\over (k_{0}-{\cal E})(k_{0}+{\cal E}^{*})}.\end{equation}
The exact advanced propagator is $D_{A}'(k)=D_{A}'(-k)$ and has poles in the first
quadrant at $k_{0}={\cal E}^{*}$ and in the second quadrant at $k_{0}=-{\cal
E}$. Consequently the free advanced quasiparticle propagator is
\begin{equation}
D_{A}(k)={1\over (k_{0}+{\cal E})(k_{0}-{\cal E}^{*})}.\end{equation}

It is straightforward to reorganize the Feynman rules for perturbation theory
so as to use the free quasiparticle propagators. 
The $2\times 2$ propagator matrix is 
\begin{eqnarray}
D_{ab}(k)=&&\left(\begin{array}{cc} D_{R}(k) & 0\cr 0 & -D_{A}(k)
\end{array}\right)\label{Dab}\\
+&& [D_{R}(k)-D_{A}(k)]n(k_{0})
\left(\begin{array}{cc} e^{\beta k_{0}} & e^{\sigma k_{0}}\cr
e^{(\beta-\sigma)k_{0}} & 1\end{array}\right),
\nonumber\end{eqnarray}
where $n(k_{0})=1/[\exp(\beta k_{0})-1]$ is the Bose-Einstein function without
absolute value bars. This satisfies the KMS condition \cite{AW1}.
Whenever there is a self-energy insertion on a propagator (i.e. $D\,\Pi\, D$)
then a counterterm $\delta \Pi$ should be inserted where
\begin{eqnarray}
\delta \Pi_{ab}(k)=&&(\vec{k}^{2}+m^{2}-{\cal E^{*}E})
\left(\begin{array}{cc} 1 & 0 \cr 0 & -1 \end{array}\right)\\
&&+ik_{0}\Gamma n(k_{0})\left(\begin{array}{cc}
e^{\beta k_{0}}+1 & -2e^{\sigma k_{0}}\cr
-2e^{(\beta-\sigma)k_{0}}  & e^{\beta k_{0}}+1\end{array}
\right).\nonumber\end{eqnarray}
The check of this counterterm is as follows. 
The inverse of Eq. (8) is
\begin{eqnarray}
D^{-1}_{ab}(k)=&&(k_{0}^{2}-{\cal E}^{*}{\cal E})
\left(\begin{array}{cc}
 1 & 0\cr
0 & -1\end{array}\right)\nonumber\\
+&&ik_{0}\Gamma n(k_{0})\left(\begin{array}{cc}
e^{\beta k_{0}}+1 & -2e^{\sigma k_{0}}\cr
-2e^{(\beta -\sigma)k_{0}}  & e^{\beta k_{0}}+1 \end{array}\right).
\end{eqnarray}
The identity
\begin{displaymath} D^{-1}_{ab}(k)-\delta\Pi_{ab}(k)
=\left(\begin{array}{cc} k^{2}-m^{2} & 0\cr
0 & -k^{2}+m^{2}\end{array}\right)\end{displaymath}
guarantees that the two approaches are identical when summed to all orders.

\subsection{One-loop retarded self-energy}

Using the quasiparticle propagator (\ref{Dab}) it is straightforward to
compute the one-loop self-energy in Fig. 1. The retarded self-energy  reduces to
\begin{eqnarray}\label{two}
\Pi_{R}(k)={g^{2}\over 2}\!\int\!&&{d^{3}k_{1}d^{3}k_{2}\over
2E_{1}2E_{2}}\delta^{3}(\vec{k}_{1}+\vec{k}_{2}-\vec{k})/(2\pi)^{3}\\
\times\Bigg\{&&{[1+n({\cal E}_{1})][1+n({\cal E}_{2})]-n({\cal E}_{1})n({\cal
E}_{2})\over k_{0}-{\cal E}_{1}-{\cal E}_{2}}\nonumber\\
+&&{[1+n({\cal E}_{1})]n({\cal E}_{2}^{*})-n({\cal E}_{1})[1+n({\cal
E}_{2}^{*})]\over k_{0}-{\cal E}_{1}+{\cal E}_{2}^{*}}\nonumber\\
+&&{n({\cal E}_{1}^{*})[1+n({\cal E}_{2})]-[1+n({\cal E}_{1}^{*})]n({\cal
E}_{2})\over k_{0}+{\cal E}_{1}^{*}-{\cal E}_{2}}\nonumber\\
+&&{n({\cal E}_{1}^{*})n({\cal E}_{2}^{*})-[1+n({\cal E}_{1}^{*})]
[1+n({\cal E}_{2}^{*})]\over k_{0}+{\cal
E}_{1}^{*}+{\cal E}_{2}^{*}}\Bigg\}.\nonumber
\end{eqnarray}
The statistical factors in the numerators of  Eq. (\ref{two}) describe 
the  emission and absorption of quasiparticles, weighted by 
$1+n({\cal E})$ and
$n({\cal E})$ respectively, and also the emission and absorption of
quasiholes,
weighted by $1+n({\cal E}^{*})$ and $n({\cal E}^{*})$.  Each of the
numerators  describes a direct process minus the inverse process. The
denominators are singular in the lower half-plane at the four locations
\begin{equation}
k_{0}=\lambda(\vec{k}_{1})+\lambda(\vec{k}_{2})\hskip
1.5cm \lambda={\cal E}\;{\rm or}\; -{\cal E}^{*}.\end{equation}
All of these lie in the lower half-plane since ${\cal E}$ and $-{\cal E}^{*}$
have the same negative imaginary part. (There are also unphysical cuts related to
Matsubara frequencies but these are cancelled by
the two-loop contribution \cite{AW1}.) Eq. (11) can also be written as a dispersion
integral over a  spectral function as discussed in Eq. (3.124) of [10].

\subsection{Two-loop retarded self-energy}

The two-loop self-energy shown in Fig. 2 requires the counterterm matrix.
 The retarded part part of the self-energy has both two-particle and
three-particle cuts. The two-particle cuts are not singular (because the
integrands contain no double poles) and have the same form as Eq. (\ref{two})
 only multiplied by wave-function renormalization factors. There is no
breakdown of perturbation theory near the production thresholds because the
location of the thresholds is now correct.  The three particle cuts of Fig. 2
occur at
\begin{equation}
k_{0}=\lambda(\vec{k}_{2})+\lambda(\vec{k}_{3})+\lambda(\vec{k}_{4}).
\end{equation}
There are eight possibilities since $\lambda={\cal E}$ or $-{\cal E}^{*}$,
independently.

The two-loop retarded self-energy shown in Fig. 3 has a variety of branch cuts.
It has two-particle cuts at
\begin{eqnarray}
k_{0}&&=\lambda(\vec{k}_{1})+\lambda(\vec{k}_{2})\nonumber\\
k_{0}&&=\lambda(\vec{k}_{3})+\lambda(\vec{k}_{4}),\end{eqnarray}
and three-particle cuts at
\begin{eqnarray}
k_{0}&&=\lambda(\vec{k}_{1})+\lambda(\vec{k}_{3})+\lambda(\vec{k}_{5})\nonumber\\
k_{0}&&=\lambda(\vec{k}_{2})+\lambda(\vec{k}_{4})+\lambda(\vec{k}_{5}).
\end{eqnarray}
A complete account of the two-loop calculations is given in \cite{AW1}.
\begin{figure}
\begin{picture}(200,110)(-10,0)
\put(100,50){\oval(100,50)}
\put(100,25){\line(0,1){50}}
\put(0,50){\line(1,0){50}}
\put(150,50){\line(1,0){50}}
\put(50,75){1}
\put(50,20){2}
\put(140,20){3}
\put(140,75){4}
\put(105,45){5}
\put(-20,0){	Fig 3: Two-loop self-energy due to vertex correction.}
\end{picture}\end{figure}

\section{Field theory of quasiparticles}

Reorganizing the Feynman rules so as to use the free quasiparticle propagator
(\ref{Dab})  accomplishes two things. First, the perturbative expansion does not
break down near the branch points. Second, the branch points of individual
diagrams are the correct branch points of the full theory. 
The original motivations given in Sec.I can now be restate more specifically:
\begin{itemize}
\item Do the quasiparticle Feynman rules come from an operator theory with
a Hamiltonian that acts in a linear vector space?
\item Are the  poles in the retarded propagator at $k_{0}={\cal E}(\vec{k})$ and
at $k_{0}=-{\cal E}(\vec{k})^{*}$   eigenvalues of a
Hamiltonian?
\item Are the energies of two-particle branch cuts at
$k_{0}=\lambda(\vec{k}_{1})+\lambda(\vec{k}_{2})$, where $\lambda(\vec{k})
={\cal E}(\vec{k})$ or $-{\cal E}^{*}(\vec{K})$, eigenvalues of
states containing two quasiparticles?
\end{itemize} 
The answer to all these questions is ``yes". However since the energies
$\lambda(\vec{k})$ are temperature-dependent
and complex the relevant  Hamiltonian 
cannot be the usual one. 

Previous treatments  have not
used the complex energy pole to define quasiparticles, but instead have used
a continuous mass spectrum
\cite{Lan,PH,BB}. These approaches are not closely related to methods for
calculation as it is unclear how to incorporate interactions.

The treatment of quasiparticles described below is based on a Lagrangian
$\check{L}$ that enjoys two properties. First, it  generates the
conventional finite-temperature perturbation series with free propagator poles
at $k^{2}=m^{2}$.  Second, by adding and subtracting the same terms  the
Lagrangian is written as  $\check{L}=\check{L}_{0}+\check{L}_{I}$ in which
$\check{L}_{0}$  describes free quasiparticles with the propagator (\ref{Dab})
and $\check{L}_{I}$ generates  the counterterm $\delta\Pi$  and the self
interactions.   

\subsection{Lagrangian for free quasiparticles}

The conjecture for the free quasiparticle Lagrangian is based on the
fact that the matrix propagator (\ref{Dab}) has poles at four distinct locations:
${\cal E}, -{\cal E}^{*}, -{\cal E}, {\cal E}^{*}$. 
 A field equation that is second order in the time derivative
can only contain two of these energies. If the fields are 
  self-adjoint,  the only possible self-adjoint
equations are
\begin{eqnarray}
&&\big(i{\partial\over\partial t}+{\cal E}\big)
\big(-i{\partial\over\partial t}+{\cal E}^{*}\big)
\phi_{R}(t,\vec{k})=0\nonumber\\ 
&&\big(i{\partial\over\partial t}+{\cal E}^{*}\big)
\big(-i{\partial\over\partial t}+{\cal E}\big)
\phi_{A}(t,\vec{k})=0.\end{eqnarray}
The Lagrangian that produces these equations is
\begin{displaymath}
\check{L}_{00}\!=\!\int \!{d{3}k\over (2\pi)^{3}}
\big[(-i{\partial\over\partial t}+{\cal E}^{*})\phi_{R}(t,-\vec{k})\big]
\big[(i{\partial\over\partial t}-{\cal E})\phi_{A}(t,\vec{k})\big].
\end{displaymath}
To convert the integration over $\vec{k}$ to an integration over
spatial coordinates requires introducing 
the spatial Fourier transform of the width $\Gamma(\vec{k})$ given in (5),
\begin{equation}\Gamma(\vec{x}-\vec{x}')=\int{d^{3}k\over (2\pi)^{3}}
e^{i\vec{k}\cdot(\vec{x}-\vec{x}')}\,\Gamma(\vec{k})\label{gamma}
\end{equation}
and similarly for ${\cal E}^{*}{\cal E}(\vec{k})$.
The Lagrangian  becomes
\begin{eqnarray}
\check{L}_{00}(t)&&=\int
d^{3}x\;\dot{\phi}_{A}(x)\dot{\phi}_{R}(x)\nonumber\\
 +&&\int
d^{3}xd^{3}x'\;\big[-\phi_{A}(x) {\cal E}^{*}{\cal
E}(\vec{x}-\vec{x}')\phi_{R}(x')\label{lfree}\\
-{1\over 2}&&\dot{\phi}_{A}(x)\Gamma(\vec{x}-\vec{x}')\phi_{R}(x')
+{1\over 2}\phi_{A}(x)\Gamma(\vec{x}-\vec{x}')\dot{\phi}_{R}(x')\big].
\nonumber\end{eqnarray}
The fields are all evaluated at  time $t=x_{0}=x_{0}'$.

Before examining how to quantize $\check{L}_{00}$,
it is helpful to present an alternative argument that provides the relation
between the fields $\phi_{R/A}$ and the original fields.
The free quasiparticle action may be written as
\begin{displaymath}
\int\! dt\,\check{L}_{00}\!=\!\int\!
{d^{4}k\over (2\pi)^{4}}\;
\phi_{R}(-k)(k_{0}+{\cal E}^{*})(k_{0}-{\cal E})
\phi_{A}(k).\end{displaymath} 
This form is closely related to the
results of Aurenche and Becherrawy \cite{AurBec}
and of van Eijck, Kobes,  and van Weert \cite{RanCh1,RanCh2}
 on transforming the matrix
propagator (\ref{Dab})  to a skew-diagonal form.   In
the notation of
\cite{RanCh2} the propagator may be `diagonalized' as 
\begin{equation}
D_{ab}(k)=V^{T}_{a\beta}(-k_{0})
\left(\begin{array}{cc} 0 &D_{A}(k) \cr
D_{R}(k)  & 0\end{array}\right)_{\beta\gamma}
V_{\gamma b}(k_{0}).\label{dmatrix}\end{equation}
Latin letters $a,b,c\dots$  denote the original $\pm$ components; Greek letters
$\alpha,\beta,\gamma,\dots$ denote $R/A$. The matrix $D_{\beta\gamma}$
has diagonal entries $D_{RR}=D_{AA}=0$ and off-diagonal entries
$D_{RA}\equiv D_{A}$, $D_{AR}\equiv D_{R}$.
The matrix $V_{\beta b}(k_{0})$
has rows labelled by $R/A$ and columns labelled by $\pm$. 
The inverse of the
center matrix $D_{\beta\gamma}(k)$ in Eq. (19) is 
\begin{displaymath}
D^{-1}_{\alpha\beta}(k)=\left[\begin{array}{cc}
0 & (k_{0}-{\cal E})(k_{0}+{\cal E}^{*})\cr
(k_{0}-{\cal E}^{*})(k_{0}+{\cal E})&0\end{array}\right].\end{displaymath}
This is a Hermitian matrix. 
Using the doublet notation
\begin{displaymath}
\phi_{\alpha}=\left(\begin{array}{c} \phi_{R}\cr \phi_{A}\end{array}
\right),\end{displaymath}
the free quasiparticle action  can be expressed as
\begin{equation}
\int\! dt\,\check{L}_{00}={1\over 2}\int{d^{4}k\over (2\pi)^{4}}\,
\phi_{\alpha}(-k)\hat{D}^{-1}_{\alpha\beta}(k)\phi_{\beta}(k)\\
.\nonumber\end{equation}
This formulation of the action suggests  
 that the usual fields $\phi_{b}$ ($b=\pm$)
 be related to the new fields $\phi_{\gamma}$ ($\gamma=R/A$) by
the matrix transformation
\begin{equation}
\phi_{b}(k)\sim \phi_{\gamma b}(k)V_{\gamma}(k_{0}).\end{equation}
The transformation is non-local in time,
\begin{equation}
\phi_{b}(t,\vec{x})\sim\int_{-\infty}^{\infty} dt'\,
\phi_{\gamma}(t+t',\vec{x})V_{\gamma b}(t').\end{equation}
and is certainly not a canonical transformation.
The symbol $\sim$ does not imply mathematical equality. However it will suggest
the full Lagrangian (26) below.

\subsection{Full Lagrangian}

The full Lagrangian will determine the evolution of the fields $\phi_{R/A}$.
It is helpful to examine the contour Lagrangian used in the
conventional path integral formulation of real-time, finite-temperature field
theory:
\begin{eqnarray}
{\cal L}^{c}=&&
{1\over 2}(\partial_{\mu}\phi_{+})(\partial^{\mu}\phi_{+})
-{1\over 2}m^{2}\phi_{+}^{2}-{g\over 3!}\phi_{+}^{3}\nonumber\\
-&&{1\over 2}(\partial_{\mu}\phi_{-})(\partial^{\mu}\phi_{-})
+{1\over 2}m^{2}\phi_{-}^{2}+{g\over 3!}\phi_{-}^{3}.\nonumber\end{eqnarray}
 Under the transformation (22)  the quadratic
terms of ${\cal L}^{c}$ become 
\begin{displaymath}
(\partial_{\mu}\phi_{R})(\partial^{\mu}\phi_{A})
-m^{2}\phi_{R}\phi_{A}.\end{displaymath}
However under (22) the cubic powers $(\phi_{\pm})^{3}$  contain products of
the fields $\phi_{R/A}$ at different times.  To avoid this it is necessary to
make a truncated Taylor series expansion
\begin{equation}
\phi_{\pm}(x)=\sum_{\ell=0}^{N-1}{\partial^{\ell}\phi_{\gamma}(x)\over 
\partial x_{0}^{\ell}}C^{\ell}_{\gamma\pm},\end{equation}
where the coefficients are temporal moments:
\begin{equation}
C^{\ell}_{\gamma
b}=\int_{-\infty}^{\infty}dt'\,{t'^{\ell}\over\ell !}
V_{\gamma b}(t')/\Gamma(1+a\ell).\end{equation} 
(The parameter $a$ guarantees convergence later. It has positive real part. At the
end of any calculation $a\to 0$.) With the definition (23), cubic  powers of
$\phi_{\pm}$  will only involve products of  fields evaluated at the same time. 
Calculations must be done at arbitrary N and subsequently the limit
N$\to\!\infty$ is performed. In this limit the moments have the the property 
\begin{equation}
\lim_{a\to 0}\sum_{\ell=0}^{\infty}(ik_{0})^{\ell}C^{\ell}_{\gamma b}
=V_{\gamma b}(k_{0}),\end{equation} which will be useful later.

When  the definition (23) is  substituted into ${\cal L}^{c}$
the Lagrangian will contain time derivatives up to order N-1. Consequently
there will be N-1 canonical momenta. To construct the Hamiltonian requires
inverting the relation between the highest time derivative of the field and the
highest momentum. This inversion is not tractable if the highest time derivative
is buried in the series definition of $\phi_{\pm}$. The solution is to
add a quadratic term containing a  time derivative of order N. The
full Lagrangian density is then
\begin{eqnarray}
\check{\cal L}&=&(\partial_{\mu}\phi_{R})(\partial^{\mu}\phi_{A})
-m^{2}\phi_{R}\phi_{A}
-h{\partial^{N}\phi_{R}\over\partial x_{0}^{N}}
{\partial^{N}\phi_{A}\over\partial x_{0}^{N}}\nonumber\\
&&-{g\over 3!}(\phi_{+})^{3}+{g\over 3!}(\phi_{-})^{3}.\end{eqnarray}
Because of the auxiliary term it is trivial to 
invert the relation between the highest time derivative and the N'th canonical
momentum.
In order that $h$ has no physical effects it will have to approach zero as 
N$\to\!\infty$. The specific choice is
\begin{equation}
h=h_{0}N^{-N}.\end{equation}

\subsection{Comments on quantization of $\check{L}$}

The Lagrangian density (26) has all time derivatives up through order N.
The construction of a Hamiltonian for such systems was developed by the
mathematician Ostrogradski in 1850 and is discussed in the mechanics book
by E.T. Whittaker \cite{Whi}. For the full Lagrangian density the 2N canonical
coordinates are the fields $\phi_{R/A}$ and their time derivatives up through
order N-1. There are 2N canonical momenta. The classical Poisson bracket
is replaced  by the quantum mechanical commutator in order to quantize.

To compute in perturbation theory it is necessary to choose the unperturbed
Lagrangian. It must include the term proportional to $h$ in order for all the
canonical momenta to appear in the unperturbed theory.
 One  choice for the unperturbed Lagrangian density would be  the three
quadratic terms on the first line of Eq. (26).  This choice ultimately leads,
in the limit  N$\to\!\infty$, to  conventional Feynman rules with propagators
containing poles at $k^{2}=m^{2}$ as shown in \cite{AW2}. 

For the unperturbed operator theory to describe quasiparticles, a different
choice of unperturbed  Lagrangian is necessary: 
\begin{equation}
\check{L}_{0}=\check{L}_{00}-h{\partial^{N}\phi_{R}\over\partial x_{0}^{N}}
{\partial^{N}\phi_{A}\over\partial x_{0}^{N}}.\end{equation}
This also has 2N canonical coordinates and 2N canonical momenta. In the limit
N$\to\!\infty$ the free propagators will be $D_{R}$ and $D_{A}$.
 The full Lagrangian (26) is separated into
$\check{L}=\check{L}_{0}+\check{L}_{I}$,
where the interaction term is
\begin{eqnarray}
\check{L}_{I}=&&\!\int\!
d^{3}x\big[\!-\vec{\nabla}\phi_{A}\cdot\vec{\nabla}\phi_{R}
-m^{2}\phi_{A}\phi_{R}-{g\over 3!}\phi_{+}^{3}+{g\over 3!}\phi_{-}^{3}\big]
\nonumber\\
+&&\int d^{3}xd^{3}x'\;\Big[\phi_{A}(x)
{\cal E}^{*}{\cal E}(\vec{x}-\vec{x}')\phi_{R}(x')\\
+{1\over 2}&&\dot{\phi}_{A}(x)\Gamma(\vec{x}-\vec{x}')\phi_{R}(x')
-{1\over 2}\phi_{A}(x)\Gamma(\vec{x}-\vec{x}')\dot{\phi}_{R}(x')\Big].
\nonumber\end{eqnarray}
The quadratic terms in $\check{L}_{I}$ will produce the counter
term matrix (9) and the cubic terms produce the interaction vertices.

\section{ Quantization of $\check{L}_{00}$}

Although the unperturbed Lagrangian $\check{L}_{0}$ 
produces field equations with time derivatives of order 2N, 
the vanishing of $h$ according to Eq. (27) is chosen so
that as N$\to\infty$ the extra modes become infinitely heavy and decouple
\cite{AW2}. 
Consequently the essential physics of the free quasiparticles
is contained in $\check{L}_{00}$. It
it therefore quite instructive to quantize $\check{L}_{00}$ first. 
This will lead to a self adjoint Hamiltonian $\check{H}_{00}$ whose
eigenvalues are the locations of the poles and cuts in the exact
thermal propagators. 

\subsection{Field operators}

The field equations that follow from $\check{L}_{00}$ are given in Eq. (16).
The solutions  can be expanded in terms of spatial Fourier transforms in a
box of volume $V$ with periodic boundary conditions:    
\begin{eqnarray}
 \phi_{R}(t,\vec{x})
&=&\!\sum_{\vec{k}}{1\over\sqrt{2EV}}
\big(a^{R}_{\vec{k}}\,e^{i\vec{k}\cdot\vec{x}-i{\cal E}^{*}t}
+a^{R\dagger}_{\vec{k}}e^{-i\vec{k}\cdot\vec{x}+i{\cal E}t}\bigr)\nonumber\\
\label{phir}\\
\phi_{A}(t,\vec{x})
&=&\!\sum_{\vec{k}}{1\over\sqrt{2EV}}
\big(a^{A}_{\vec{k}}\,e^{i\vec{k}\cdot\vec{x}-i{\cal E}t}
+a^{A\dagger}_{\vec{k}}\,e^{-i\vec{k}\cdot\vec{x}+i{\cal
E}^{*}t}\bigr).\nonumber\end{eqnarray}
Note that $\phi_{R}(t,\vec{k})\!\sim\exp(+\Gamma t/2)$ rises exponentially with
time and  $\phi_{A}(t,\vec{k})\!\sim\exp(-\Gamma t/2)$ falls exponentially.
The Hamiltonian will contain products of the two fields and will be
time-independent.  For a fixed wave vector $\vec{k}$, one may interpret
$\phi_{A}(t,\vec{k})$ 
and $\phi_{R}(t,\vec{k})$ as the coordinates of a damped harmonic oscillator
an  anti-damped harmonic oscillator, respectively. The oscillator problem, devoid
of the field concept, has a long history. Bateman \cite{Bate} in 1931 invented a
Lagrangian of the form (18) to treat  the classical problem of one
damped and one anti-damped oscillator. Morse and Feshbach \cite{MorFes} also use
this same oscillator Lagrangian. The quantum mechanics of the system is
discussed in \cite{Fes,Dek}.  

\subsection{Canonical momenta and commutation relations}

Since there are only first order time derivatives, the usual 
canonical formalism applies directly. 
The conjugate momenta are $\pi_{\alpha}=\delta\check{L}_{00}/\delta
\dot{\phi}_{\alpha}$:
\begin{eqnarray}
\pi_{R}(x)&=&\dot{\phi}_{A}(x)+
{1\over 2}\int
d^{3}x'\,\Gamma(\vec{x}-\vec{x}')\phi_{A}(x')\nonumber\\
\pi_{A}(x)&=&\dot{\phi}_{R}(x)-{1\over 2} \int
d^{3}x'\,\Gamma(\vec{x}-\vec{x}')\phi_{R}(x').
\nonumber\end{eqnarray}
 The canonical commutation relations 
are 
\begin{equation}
\Big[\pi_{\alpha}(x),\phi_{\beta}(y)\Big]_{x_{0}=y_{0}}
={1\over i}\delta_{\alpha\beta}\,\delta^{3}(\vec{x}-\vec{y}).\end{equation} 
 Consequently the raising and lowering operators  satisfy
\begin{equation}
\Big[a^{R}_{\vec{k}},a^{A\dagger}_{\vec{k}'}\Big]
=\Big[a^{A}_{\vec{k}},a^{R\dagger}_{\vec{k}'}\Big]=\delta_{\vec{k},\vec{k}'}
.\end{equation}
The other commutators vanish: $[a_{\vec{k}}^{R},a_{\vec{p}}^{R\dagger}]
=[a_{\vec{k}}^{A},a_{\vec{p}}^{A\dagger}]=0$ and
$[a_{\vec{k}}^{R},a_{\vec{p}}^{A}]=0$.
This implies $[\phi_{R}(x),\phi_{R}(y)]=0$ and $[\phi_{A}(x),\phi_{A}(y)]=0$.
Because the time dependence of $\phi_{R}$ and $\phi_{A}$,
 if these two field commutators  
did not vanish they would have to  depend on both  $x_{0}$ and $y_{0}$ 
(rather than $x_{0}-y_{0}$) and would thus violate invariance under time
translations.
The only non-vanishing commutator of two fields is
$[\phi_{A}(x),\phi_{R}(y)]$ and it automatically satisfies translational
invariance.

\subsection{Hamiltonian $\check{H}_{00}$}

The Hamiltonian 
$\check{H}_{00}=\int
d^{3}x\,\Big[\pi^{R}\dot{\phi}_{R}+\pi^{A}\dot{\phi}_{A}\big]-\check{L}_{00}$
describes free quasiparticles. In terms of
canonical variables it is
\begin{eqnarray}
\check{H}_{00}=\int
d^{3}xd^{3}x'\big[&&\pi_{A}(x)\delta(\vec{x}-\vec{x}')\pi_{R}(x')\nonumber\\
-&&{1\over2}\pi_{A}(x)\Gamma(\vec{x}-\vec{x}')\phi_{A}(x')\nonumber\\
+&&{1\over 2}\phi_{R}(x)\Gamma(\vec{x}-\vec{x}')\pi_{R}(x')\nonumber\\
+&&\phi_{R}(x)E^{2}(\vec{x}-\vec{x}')\phi_{A}(x')\Big].\nonumber
\end{eqnarray}
The  kernel $E^{2}(\vec{x}-\vec{x}')$ is the Fourier transform of
$E^{2}(\vec{k})$ analogous to (17).
When the plane wave expansions (30) are substituted
  the Hamiltonian reduces to
\begin{equation}
\check{H}_{00}=\sum_{\vec{k}}\big[\;{\cal
E}\,(a^{R\dagger}_{\vec{k}}a^{A}_{\vec{k}} +{1\over 2})
+{\cal E}^{*}\,(a^{A\dagger}_{\vec{k}}a^{R}_{\vec{k}}+{1\over
2})\;\big],\end{equation}
which is manifestly self-adjoint.
The complete analysis of this Hamiltonian is rather lengthly. 
The results of that analysis will be summarized here but not proven.
The essential feature is that there are two parallel sets of eigenstates. One set
is built upon a ground state $|0_{R}\rangle$; the other, on a ground state
$|0_{A}\rangle$. The retarded
vacuum state $|0_{R}\rangle$ is defined by the conditions
\begin{equation}
a^{A}_{\vec{k}}|0_{R}\rangle=0
\hskip1cm
a^{A\dagger}_{\vec{k}}|0_{R}\rangle=0.
\end{equation}
(Recall that $a^{A}$ and $a^{A\dagger}$ commute.)
The retarded vacuum is an eigenstate of $\check{H}_{00}$. (Since
$\check{H}_{00}$ is not normal ordered, the eigenvalue is not zero.) The
simplest excitations of the vacuum are the addition of one quasiparticle,
$a^{R\dagger}_{\vec{k}}|0_{R}\rangle$, or the addition of one quasihole,
$a^{R}_{\vec{k}}|0_{R}\rangle$.
From the commutation relations (32) these operators shift the energy by 
${\cal E}$ or $-{\cal E}^{*}$, respectively:
\begin{displaymath}
\big[\check{H}_{00},a^{R\dagger}_{\vec{k}}\big]={\cal
E}a^{R\dagger}_{\vec{k}}
\hskip1cm
\big[\check{H}_{00},a^{R}_{\vec{k}}\big]=-{\cal E}^{*}a^{R}
_{\vec{k}}.\end{displaymath}
Note that both ${\cal E}$ and $-{\cal E}^{*}$ have the same negative imaginary
part. The next excitations are those with two quasiparticles,
$a^{R\dagger}a^{R\dagger}|0_{R}\rangle$, with one quasiparticle and one
quasihole, $a^{R\dagger}a^{R}|0_{R}\rangle$, or with
two quasiholes, $a^{R}a^{R}|0_{R}\rangle$.  Every state of the form
$(a^{R\dagger})^{\ell}(a^{R})^{m}|0_{R}\rangle$ is of the retarded type and
has an energy with a negative imaginary part. (The real part of the energy may
be of either sign.)

There is another set of states that are excitations of the advanced
vacuum state $|0_{A}\rangle$, which is defined by the conditions
\begin{equation}
a^{R}_{\vec{k}}|0_{A}\rangle=0
\hskip1cm
a^{R\dagger}_{\vec{k}}|0_{A}\rangle=0.
\end{equation}
 The advanced representation of the one quasiparticle excitation is
$a^{A\dagger}_{\vec{k}}|0_{A}\rangle$ and of  the  one quasihole excitation,
$a^{A}_{\vec{k}}|0_{A}\rangle$.
From the commutation relations (32) these operators shift the energy by 
${\cal E}^{*}$ or $-{\cal E}$, respectively:
\begin{displaymath}
\big[\check{H}_{00},a^{A\dagger}_{\vec{k}}\big]={\cal
E}^{*}a^{A\dagger}_{\vec{k}}
\hskip1cm
\big[\check{H}_{00},a^{A}_{\vec{k}}\big]=-{\cal E}a^{A}
_{\vec{k}}.\end{displaymath}
Both ${\cal E}^{*}$ and $-{\cal E}$ have the same positive imaginary
part.  The advanced representation  of a general excitation is
$(a^{A\dagger})^{\ell}(a^{A})^{m}|0_{A}\rangle$ and
has an energy with a positive imaginary part. 

As explained in Appendix A, the Hamiltonian does not act in a Hilbert space. None
of the retarded states
$|\Psi_{R}\rangle$ have finite norm; none of the advanced states
$|\Psi_{A}\rangle$ have finite norm. However the scalar product between retarded
and advanced states,
$\langle \Psi'_{A}|\Psi_{R}\rangle$, is always finite.
For example, $\langle
0_{A}|a^{A}_{\vec{k}}a^{R\dagger}_{\vec{k}}|0_{R}\rangle=1$ and 
$\langle 0_{A}|a^{A\dagger}_{\vec{k}}a^{R}_{\vec{k}}|0_{R}\rangle=-1$.
 Physical applications
always involve matrix elements of this form. 
The completeness relation, 
\begin{equation}
1=\sum_{\Psi}|\Psi_{R}\rangle\,\langle\Psi_{A}|,\end{equation}
is useful in evaluating matrix elements of the form
$\langle 0_{A}|XY|0_{R}\rangle$ as shown in the next section.

\section{Sample results}

It is not possible in the allotted space to give a full discussion of this
formalism. One of the important features that will be used below is that
all the temperature-dependence is contained in the full Lagrangian (26).
Consequently propagators, Green functions, and the free energy are computed as
vacuum matrix elements; there are no thermal traces.

\subsection{Free propagators for $\phi_{R/A}$}

The quasiparticle pole in the propagator  comes directly from
the eigenstate of the Hamiltonian $\check{H}_{00}$. The free retarded propagator
is defined as a time-ordered product:
\begin{eqnarray}
D_{AR}(x)=&&-i\langle 0_{A}|T\big(\phi_{A}(x)\phi_{R}(0)\big)|0_{R}\rangle
\nonumber\\
=&&-i\theta(x_{0})\langle
0_{A}|\phi_{A}(x)\phi_{R}(0)|0_{R}\rangle.\end{eqnarray}
Only one time-ordering survives on the second line because
 $\langle 0_{A}|\phi_{R}(0)\phi_{A}(x)|0_{R}\rangle=0$.
When  the plane wave solutions (30) are inserted, the result is
\begin{displaymath}
D_{AR}(x)=-i\theta(t)\sum_{\vec{k}}
{1\over 2EV}\Big(e^{i\vec{k}\cdot\vec{x}-i{\cal E}t}
-e^{-i\vec{k}\cdot\vec{x}+i{\cal E}^{*}t}\Big).\end{displaymath}
The four-dimensional Fourier transform is
\begin{equation}
D_{AR}(k)={1\over (k_{0}-{\cal E})(k_{0}+{\cal E}^{*})}.\end{equation}
Thus the operator theory produces the quasiparticle propagator as desired.
Similarly, the free advanced quasiparticle propagator is
\begin{eqnarray}
D_{RA}(x)=&&-i\langle 0_{A}|T\big(\phi_{R}(x)\phi_{A}(0)\big)|0_{R}\rangle
\nonumber\\
=&&-i\theta(-x_{0})\langle
0_{A}|\phi_{A}(0)\phi_{R}(x)|0_{R}\rangle\end{eqnarray}
Comparison with (37) shows that $D_{RA}(x)=D_{AR}(-x)$.
Thus $D_{RA}(k)=D_{RA}(-k)$.
Since $\langle 0_{A}|\phi_{R}\phi_{R}|0_{R}\rangle=0$ and 
$\langle 0_{A}|\phi_{A}\phi_{A}|0_{R}\rangle=0$ the matrix propagator in the
retarded/advanced basis is
\begin{equation}
D_{\alpha\beta}(k)=\left(\begin{array}{cc}
0 & D_{RA}(k)\\ D_{AR}(k) & 0 \end{array}\right)\end{equation}
in agreement with Eq. (19).

\subsection{Free propagators for $\phi_{\pm}$}

Although the complex quasiparticle energy ${\cal E}(\vec{k})$  is
temperature-dependent, there are no  Bose-Einstein functions in the
retarded or advanced propagators (40). However, the  Bose-Einstein functions 
should appear in the propagators for $\phi_{\pm}$ in order to agree with Eq.
(\ref{Dab}). The
$\phi_{\pm}$ propagators are vacuum matrix elements:
\begin{equation}
G_{ab}(x-y)=-i\langle 0_{A}|T\Big(\phi_{a}(x)\phi_{b}(y)\Big)|0_{R}\rangle,
 \end{equation}
where $a,b=\pm$.
To compute this properly requires quantizing not $\check{L}_{00}$ but
$\check{L}_{0}$ given in Eq. (28). I will provide the essential feature of that
quantization at the appropriate point below. 

Using the definition (23) of the $\phi_{\pm}$ as a sum of the first
N-1 time derivatives of $\phi_{R}$ and $\phi_{A}$ gives for $G_{ab}$
\begin{displaymath}
G_{ab}(x-y)=\!-i\!\sum_{\ell,m=0}^{N-1}\langle 0_{A}|T\Big(
{\partial \phi_{\beta}(x)\over\partial x_{0}^{\ell}}
{\partial \phi_{\gamma}(y)\over\partial y_{0}^{m}}\Big)|0_{R}\rangle
C^{\ell}_{\beta a}C^{m}_{\gamma b}.\end{displaymath}
When $\phi_{R/A}$ are quantized according to $\check{L}_{0}$ the time
derivatives may be taken outside of the time-ordering so that
\begin{displaymath}
G_{ab}(x-y)=\sum_{\ell,m=0}^{N-1}
{\partial \over\partial x_{0}^{\ell}}
{\partial \over\partial y_{0}^{m}}G_{\beta\gamma}(x-y)
C^{\ell}_{\beta a}C^{m}_{\gamma b}.\end{displaymath}
The Fourier transform is
\begin{displaymath}
G_{ab}(k)=\sum_{\ell=0}^{N-1}(-ik_{0})^{\ell}C^{\ell}_{\beta a}
\big[ G_{\beta\gamma}(k)\big]\sum_{m=0}^{N-1}(ik_{0})^{m}C^{m}_{\gamma
b}\end{displaymath} 
As the number of derivatives goes to infinity the identity (25)  allows the 
moments to be summed:
\begin{displaymath}
\lim_{N\to\infty}G_{ab}(k)\!=\!V^{T}_{a\beta}(-k_{0})\left(
\begin{array}{cc} 0 &D_{RA}(k)\cr
D_{AR}(k) & 0 \end{array}\right)_{\beta\gamma}
\!V_{\gamma b}(k_{0}).\end{displaymath}
This is exactly the result (\ref{dmatrix}) for the matrix propagator of
quasiparticles. The Bose-Einstein functions are contained in $V(k_{0})$.

\subsection{One-loop self-energy}

 There is a pleasant surprise
when we examine the interacting propagator. 
The one-loop self-energy, for example, is
\begin{equation}
\Pi_{bc}(x-y)={-ig^{2}\over 4}\sigma_{b}\sigma_{c}\langle 0_{A}|
T\big(\phi_{b}^{2}(x)\phi_{c}^{2}(y)\big)|0_{R}\rangle.
\end{equation}
One way to compute this is to use Wick's theorem to express the matrix element
as $\big[D_{bc}(x-y)\big]^{2}$. Computing $\Pi_{bc}(k)$ then requires
integrating the one-loop diagram Fig. 1 with quasiparticle propagators. The
result is already given in Eq. (11).

The operator expression (42) can be evaluated in another manner that 
 explains all the branch cuts displayed in Eq. (11).
The quasiparticle Hamiltonian determines the time-dependence of the field
operators:
\begin{displaymath}
\phi_{b}(x)=e^{i\check{H}_{00}\,x_{0}}\phi_{b}(\vec{x})e^{-i\check{H}_{00}\,x_{0}}
\end{displaymath}
as can be easily verified.
For either of the two time-orderings in Eq. (42) a complete set of states may be
inserted between the operators using
the completeness relation (36).
The states are energy eigenstates satisfying 
$\check{H}_{00}|\Psi_{R}\rangle=|\Psi_{R}\rangle {\cal E}_{\Psi}^{R}$ and
$\check{H}_{00}|\Psi_{A}\rangle=|\Psi_{A}\rangle {\cal E}_{\Psi}^{A}$
with ${\cal E}_{\Psi}^{A}={\cal E}_{\Psi}^{R*}$.
The matrix element in Eq. (42) has the value
\begin{eqnarray}
\sum_{\Psi}\!\Big(\!\theta(x_{0}\!-\!y_{0})&&\langle
0_{A}|\phi_{b}^{2}(\vec{x})|\Psi_{R} \rangle 
\langle \Psi_{A}|\phi_{c}^{2}(\vec{y})|0_{R}\rangle
e^{-i{\cal E}_{\Psi}^{R}(x_{0}-y_{0})}\nonumber\\
+\theta(y_{0}\!-\!x_{0})&&\langle
0_{A}|\phi_{c}^{2}(\vec{y})|\Psi_{R} \rangle 
\langle \Psi_{A}|\phi_{b}^{2}(\vec{x})|0_{R}\rangle
e^{i{\cal E}_{\Psi}^{R}(x_{0}-y_{0})}\!\Big).\nonumber\end{eqnarray}
Since Im ${\cal E}_{\Psi}^{R}<0$ the time dependence is exponentially falling in
both terms. The temporal Fourier transform to $k_{0}$ at fixed $\vec{x}$ and
$\vec{y}$ is 
\begin{eqnarray}
\Pi_{bc}={g^{2}\over
4}\sigma_{b}\sigma_{c}\sum_{\Psi}\Big(&&{\langle
0_{A}|\phi_{b}^{2}(\vec{x})|\Psi_{R}
\rangle 
\langle \Psi_{A}|\phi_{c}^{2}(\vec{y})|0_{R}\rangle
\over k_{0}-{\cal E}_{\Psi}^{R}}\nonumber\\
-&&{\langle 0_{A}|\phi_{c}^{2}(\vec{y})|\Psi_{R} \rangle 
\langle \Psi_{A}|\phi_{b}^{2}(\vec{x})|0_{R}\rangle
\over k_{0}+{\cal E}_{\Psi}^{R}}\Big)\end{eqnarray}
The operators $\phi_{b/c}^{2}$ only connect the vacuum with states containing
two excitations (i.e. two quasiparticles, one quasiparticle and one quasihole,
or two quasihole). Thus the possibilities are
\begin{equation}\begin{array}{rc}
|\Psi_{R}\rangle  & {\cal E}_{\Psi}^{R}\;\;\cr\cr
a^{R\dagger}_{\vec{k}_{1}}a^{R\dagger}_{\vec{k}_{2}}|0_{R}\rangle
&{\cal E}(\vec{k}_{1})\!+\!{\cal E}(\vec{k}_{2})\cr
a^{R\dagger}_{\vec{k}_{1}}a^{R}_{\vec{k}_{2}}|0_{R}\rangle
&{\cal E}(\vec{k}_{1})\!-\!{\cal E}^{*}(\vec{k}_{2})\cr
a^{R}_{\vec{k}_{1}}a^{R\dagger}_{\vec{k}_{2}}|0_{R}\rangle
&-{\cal E}^{*}(\vec{k}_{1})\!+\!{\cal E}(\vec{k}_{2})\cr
a^{R}_{\vec{k}_{1}}a^{R}_{\vec{k}_{2}}|0_{R}\rangle
&-{\cal E}^{*}(\vec{k}_{1})\!-\!{\cal E}^{*}(\vec{k}_{2})\cr
\end{array}\end{equation}
The four terms in $\Pi_{bc}$ with denominators $k_{0}-{\cal E}_{\Psi}^{R}$ all
have branch cuts in the lower-half of the complex $k_{0}$ plane.
These produce the four
branch cuts previously found  in Eq. (11). The retarded self-energy is the
particular linear combination of the $\Pi_{bc}$ which contains only these four
cuts.  The four terms in Eq. (43) with denominators $k_{0}+{\cal E}_{\Psi}^{R}$
all have branch cuts in the upper-half of the complex $k_{0}$ plane and produce
the advanced self-energy.  The numerators in Eq. (43) may be evaluated using
 relations (23) and (25) and result in the correct Bose-Einstein functions
$n({\cal E})$ and $n({\cal E}^{*})$ in agreement with Eq. (11). 

\section{Discussion}

The treatment presented here is based on the fact that
field theories at finite temperature inevitably have poles in the full
propagator at complex, temperature-dependent energies
${\cal E}(\vec{k})$. To consistently include quasiparticles in Feynman
diagram computations it is only necessary to use the quasiparticle propagator
(8) and the counter-term (9). To go further and incorporate the quasiparticle
effects into the operator structure requires the temperature-dependent Lagrangian
(26). With this Lagrangian the questions posed at the beginning of
Sec. IV may all be answered affirmatively.

The logical necessity of this development is rather tight.
As shown in Sec. II,  higher loop corrections will result in a breakdown
of perturbation theory unless the free propagators are chosen to have the
correct quasiparticle poles. The free quasiparticle  propagator shown in
Eq. (8) must then contain the off-shell Bose-Einstein function $n(k_{0})$ in
order to satisfy the KMS condition. (With quasiparticle poles, neither the
quasiparticle energy
${\cal E}(\vec{k})$ nor the T=0 energy
$E_{0}(\vec{k})=(\vec{k}^{2}+m^{2})^{1/2}$ can be the argument of the
Bose-Einstein function.) The
off-shell $n(k_{0})$  leads to the time-dependent relation (22), or the more
precise relation (23), between the original fields and $\phi_{R/A}$.  
That  fixes the Lagrangian to be (26).

The standard operator formulation of finite temperature field theory is
thermofield dynamics \cite{1,2,3}. The unperturbed propagators in thermofield
dynamics have poles at $k^{2}=m^{2}$.  The matrix propagators contain the
Bose-Einstein function with  the T=0 energy as argument: $n(E_{0})$. 
Consequently the transformation from the original fields
$\phi_{\pm}$ to the TFD fields $\phi$ and $\tilde{\phi}$ depends on space rather
than on time. The transformation is automatically canonical. See Appendix B of
[5]. (When TFD is extended to nonequilibrium situations, the transformation is
time-dependent
\cite{PH,Arim}.)

In both formulations the number of degrees of freedom is doubled.
In thermofield dynamics the operators  act in a Hilbert
space. The basis states are built out of two commuting
raising operators, $a^{\dagger}$ and $\tilde{a}^{\dagger}$.
In the approach described here, operators do not act in a Hilbert space, as
discussed in Appendix A. 
Furthermore, there are two parallel sets of basis states
$|\Psi_{R}\rangle$  and $|\Psi_{A}\rangle$. The former
are built from commuting raising operators $a^{R\dagger}$ and
$a^{R}$ acting on
$|0_{R}\rangle$; the latter from $a^{A\dagger}$ and
$a^{A}$ acting on $|0_{A}\rangle$.

\acknowledgments

It is a pleasure to thank Ulrich Heinz for hosting the Vth International
Workshop on Thermal Field Theories in Regensburg.
This work was supported in part by the U.S. National Science Foundation
under grant PHY-9630149.

\appendix

\section{Complex eigenvalues from a self-adjoint Hamiltonian}

States in a Hilbert space have finite norms and consequently self-adjoint
operators cannot have complex eigenvalues. The state space 
for quasiparticles is not a Hilbert space. A  retarded eigenstate of the
quasiparticle Hamiltonian satisfies
\begin{equation}
\check{H}_{00}|\Psi_{R}\rangle=|\Psi_{R}\rangle {\cal E}^{R}.\end{equation}
Since $\check{H}_{00}$ is self-adjoint, the adjoint of this equation is
$\langle \Psi_{R}|\check{H}_{00}={\cal E}^{R*}\langle\Psi_{R}|$.
The inner product of this with $|\Psi_{R}\rangle$ would force ${\cal E}^{R}$ to
be real If the norm $\langle\Psi_{R}|\Psi_{R}\rangle$ existed.
As shown below, this norm does not exist. 
Advanced eigenstates satisfy
$\check{H}_{00}|\Phi_{A}\rangle=|\Phi_{A}\rangle {\cal E}^{A}$. The scalar
product of this with  $\langle\Psi_{R}|$ does exist and implies 
\begin{equation}
\langle \Psi_{R}|\Phi_{A}\rangle {\cal E}^{R*}=
\langle \Psi_{R}|\Phi_{A}\rangle {\cal E}^{A}.\end{equation}
Thus if ${\cal E}^{R*}\neq {\cal E}^{A}$ then the states are orthogonal.

A concrete realization of the quasiparticle algebra is illuminating.
The four operators $a^{A}, a^{A\dagger}, a^{R}, a^{R\dagger}$ may be
represented in a two-dimensional real space as linear combinations of
$x, \partial/\partial x, y, \partial/\partial y$ as follows
\begin{eqnarray}
&& a^{A}+a^{A\dagger}=x-i{\partial\over\partial y}\hskip0.6cm
 a^{A}-a^{A\dagger}={\partial\over\partial x} +iy\nonumber\\
&&a^{R}+a^{R\dagger}=x+i{\partial\over\partial y}\hskip0.6cm
a^{R}-a^{R\dagger}={\partial\over\partial x}-iy
\end{eqnarray}
These satisfy the commutation relations (32). The free quasiparticle Hamiltonian
(33) in this representation is 
\begin{displaymath}
\check{H}_{00}={E\over 2}\big(-{\partial^{2}\over\partial x^{2}}
+x^{2}+{\partial^{2}\over\partial y^{2}}-y^{2}\big)
+{\Gamma\over 2}\big({\partial^{2}\over\partial x\partial y}+xy\big).
\end{displaymath}
The retarded vacuum state,
 defined by the conditions $a^{A}|0_{R}\rangle=
a^{A\dagger}|0_{R}\rangle=0$, has the representation
\begin{equation}
\langle\vec{x}|0_{R}\rangle=\exp(-ixy)/\sqrt{\pi},\end{equation}
which is not square integrable. The wave function is an eigenstate of
$\check{H}_{00}$ with eigenvalue $-i\Gamma/2$.
 Application of the quasiparticle creation
operator $a^{R\dagger}$ and the quasihole creation operator $a^{R}$ give
excited state wave functions 
\begin{displaymath}
\langle\vec{x}|(a^{R\dagger})^{\ell}(a^{R})^{m}|0_{R}\rangle
=H_{\ell}(x+iy)\,H_{m}(x-iy)
\,{e^{-ixy}\over\sqrt{\pi}},\end{displaymath} 
where $H_{\ell}$ and $H_{m}$ are Hermite polynomials.
None of these are square integrable. 
 These are eigenstates of
$\check{H}_{00}$ with eigenvalue 
${\cal E}(\ell+1/2)-{\cal E}^{*}(m+1/2)$. 

The advanced vacuum state
is defined by  $a^{R}|0_{A}\rangle=
a^{R\dagger}|0_{A}\rangle=0$ and has the representation
\begin{equation}
\langle\vec{x}|0_{A}\rangle=\exp(ixy)/\sqrt{\pi}.\end{equation}
It is not square integrable. However the scalar product of the advanced and
retarded vacuum states is finite:
\begin{equation}
\langle 0_{A}|0_{R}\rangle=\int_{-\infty}^{\infty}dxdy\,{\exp(-i2xy)
\over\pi}=1.\end{equation}
The excited state wave functions are
\begin{displaymath}
\langle\vec{x}|(a^{A\dagger})^{\ell}(a^{A})^{m}|0_{A}\rangle
=H_{\ell}(x-iy)\,H_{m}(x+iy)
\,{e^{ixy}\over\sqrt{\pi}}.\end{displaymath} 
All scalar products $\langle\Phi_{A}|\Psi_{R}\rangle$ are finite.


\references

\bibitem{KKR} R. Kobes, G. Kunstatter, and A. Rebhan, Phys. Rev. Lett.
and Nucl. Phys. {\bf B335}, 1 (1991).

\bibitem{BP} E. Braaten and R.D. Pisarski, Phys. Rev. Lett. {\bf 64}, 1338
(1990); Nucl. Phys. {\bf B337}, 569 (1990) and {\bf B339}, 310 (1990)..

\bibitem{lat} A. Peshier, B. Kampfer, O.P. Pavlenko, and G. Soff,
Phys. Rev. {\bf D54}, 2399 (1996); P. Levai and U. Heinz, Phys. Rev. {\bf C57},
1879 (1998); A. Peshier, these proceedings.

\bibitem{AW1} H.A. Weldon,  Phys. Rev. {\bf D58}, 105002 (1998).

\bibitem{AW2} H.A.Weldon,  Nucl. Phys. {\bf B534}, 467 (1998).

\bibitem{KS} R. Kobes and G. Semenoff, Nucl. Phys.  {\bf B260}, 714 (1985)
and {\bf B272}, 329 (1986).

\bibitem{LB} M. Le Bellac, {\it Thermal Field Theory} (Cambridge University
Press,  Cambridge, England, 1996).

\bibitem{Gel} F. Gelis, Nucl. Phys. {\bf B508}, 483 (1997).

\bibitem{Lan} N.P. Landsman, Phys. Rev. Lett. {\bf 60}, 1990 (1988) and
Ann. Phys. {\bf 186}, 141 (1988).

\bibitem{PH} P.A. Henning, Phys. Rep. {\bf 253}, 235 (1995);
P.A. Henning and H. Umezawa, Phys. Lett. {\bf B303}, 209 (1993).

\bibitem{BB} J. Bros and D. Bucholz, Z. Phys. C {\bf 55}, 509 (1992); Ann.
Inst. Henri Poincar'{e} {\bf 64}, 495 (1996).

\bibitem{AurBec} P. Aurenche and T. Becherrawy, Nucl. Phys. {\bf B379} (1992)
249.

\bibitem{RanCh1} M.A. van Eijck and Ch. G. van Weert, Phys. Lett. {\bf 278B}
(1992) 305.

\bibitem{RanCh2} M.A. van Eijck, R. Kobes, and Ch. G. van Weert, Phys. Rev.
{\bf D50} (1994) 4097.

\bibitem{Whi} E.T. Whittaker, {\it A Treatise on the Analytical Dynamics
of Particles and Rigid Bodies}, 4'th ed. (Cambridge University Press, Cambridge,
England, 1947) p. 265-267.

\bibitem{Bate} H. Bateman, Phys. Rev. {\bf 38}, 815 (1938).

\bibitem{MorFes} P. Morse and H. Feshbach, {\it Methods of Mathematical
Physics} (McGraw Hill, New York, 1953) p. 298.

\bibitem{Fes} H. Feshbach and YT. Tikochinsky, in {\it A Festschrift for I.I.
Rabi} (Trans. N.Y. Acad. Sc.) Ser. 2, Vol {\bf 38}, 44 (1977).

\bibitem{Dek} H. Dekker, Phys. Rep. {\bf 80}, 1 (1980).

\bibitem{1} H. Umezawa, Y.H. Matsumoto and M. Tachiki,{\it Thermo Field
Dynamics and Condensed States} ( North Holland, Amsterdam, Netherlands, 1982).

\bibitem{2} I. Ojima, Ann. Phys. (N.Y.) {\bf 137},1 (1981);
H. Matsumoto, I. Ojima, and H. Umezawa, Ann. Phys. (N.Y.) {\bf 152}, 348 (1984).

\bibitem{3} H. Umezawa, {\it Advanced Field Theory: Micro, Macro, and Thermal
Physics} (American Institute of Physics, New York, 1993).

\bibitem{Arim} T. Arimitsu and H. Umezawa, Prog. Th. Phys. {\bf 77}, 32 and  53
(1987); T. Arimitsu, Cond. Mat. Phys. (Ukranian Nat. Acad. Sc.) {\bf 4},
26 (1994).
\end{document}